\documentclass[vecphys]{svmult}
\usepackage{makeidx}
\usepackage{graphicx} 
\usepackage{color,epsfig,latexsym,epsf}
\usepackage{graphics}
\usepackage{rotate}
\usepackage{multicol}
\usepackage[bottom]{footmisc}

\begin{document}

\title*{REGULAR AND CHAOTIC MOTION IN ELLIPTICAL GALAXIES}

\author{JUAN C. MUZZIO}
\institute{Facultad de Ciencias Astron\'omicas y Geof\'{\i}sicas
de la UNLP and Instituto de Astrof\'{\i}sica de La Plata (UNLP--CONICET),
Observatorio Astron\'omico, 1900 La Plata, Argentina}
\authorrunning{Muzzio J.C.}
\titlerunning{Motion in elliptical galaxies}
\maketitle
\begin{abstract}
Here I review recent work, by other authors and by myself, on some
particular topics related to the regular and chaotic motion in
elliptical galaxies. I show that it is quite possible to build
highly stable triaxial stellar systems that include large
fractions of chaotic orbits and that partially and fully chaotic orbits
fill different regions of space, so that it is important not to group
them together under the single denomination of chaotic orbits. Partially
chaotic orbits should not be confused with weakly fully chaotic orbits
either, and their spatial distributions are also different. Slow figure
rotation (i.e., rotation in systems with zero angular momentum) seems to
be always present in highly flattened models that result from cold
collapses, with the rotational velocity diminishing
or becoming negligibly small for less flattened models. Finally, I
comment on the usefulness and limitations of the classification of
regular orbits via frequency analysis.
\end{abstract}
\section{Introduction}
\label{sec:1}
It is fitting, in this conference in memory of N. Voglis, to recall that
I became interested in the investigation of regular and chaotic motions
in elliptical galaxies thanks to a paper of his \cite{VKS02}. By that time,
I had been working on N--body problems for two decades, and on regular
and chaotic motion for seven or eight years, but I had never been involved
in research on elliptical galaxies. The paper by Voglis and his coworkers
showed me that, with the computers and the numerical tools I had at my
disposal, I might be able to contribute significantly to a very interesting
subject and, in fact, I have been devoted to that subject ever since.

Having worked in this field for a few years only, it would be presumtuous
from my part to attempt to present here a complete review of the subject.
Alternatively, to be a relative newcomer to the field has the advantage of
bringing to it views and opinions different from the prevailing ones: they
may be wrong, but they stimulate progress.

Therefore, I will limit the scope of this review to a few items that have
been of particular interest to me and which I have strived to clarify with
my research: 1) Can we build stable triaxial models of stellar systems
that contain high fractions of chaotic orbits?; 2) Is the distinction
between partially and fully chaotic orbits of any use?; 3) Is figure
rotation significant in triaxial stellar systems?; 4) Which are the
usefulness and limitations of frequency analysis for the classification of
large numbers of regular orbits in model stellar systems?

Since galactic dynamics is not the only subject of this conference, which
includes other fields like celestial mechanics, it may be useful to recall
that the time scales pertinent to galaxies are completely different from
those that rule the Solar System. While the age of the latter is of the
order of $10^8$ orbital periods, galactic ages are of the order of $10^2$
orbital periods only. Thus, the chaotic orbits we will refer here are much
more strongly chaotic (i.e., their Lyapunov times measured in orbital periods
are much shorter) than those of the Solar System. Technical tools, such as
frequency analysis, should also be considered with this fact in mind.

\section{Highly chaotic triaxial stellar systems}
\label{sec:2}

\subsection{Building self--consistent triaxial stellar systems}
\label{sec:3}
A popular method to build a self--consistent triaxial stellar system is the one
due to Schwarzschild \cite{Sch79}. One chooses a density distribution and
obtains the potential that it creates; a library of thousands of orbits is
then computed in that potential and weights are assigned according to the
time that a body on that orbit spends in different regions of space; 
finally, those weights are used to compute the relative numbers of those orbits
that are needed to obtain the original density distribution.

Another way to proceed is to use an N--body code to build a triaxial stellar
system (e.g., through the collapse of an N--body system initially out of
equilibrium), then to smooth and to freeze the potential fitting it with adequate
formulae, to use these formulae, together with the positions and velocities of the
bodies as initial conditions, to compute a representative sample of orbits in that
potential and, finally, to classify those orbits to get the orbital structure of
the system
\cite{VKS02}.

Those two methods should be regarded as complementary. Schwarzschild's one
allows a very precise definition of the density distribution of the system
one wants to study; alternatively, some properties of the models dictated by
mathematical simplicity (e.g., constant axial ratios over the whole system)
might bias its results, while the N--body method is free of that problem.

\subsection{The problem of chaotic orbits in Schwarzschild's method}
\label{sec:4}
Schwarzschild \cite{Sch93} found it necessary to include chaotic orbits in his
models but, then, these were not fully stable. He built several models using
orbits computed over a Hubble time and, subsequently, followed those orbits for
two additional Hubble times. When he computed the axial ratios obtained using
the data for the third Hubble time, he found significant differences with
respect to the ratios computed over the first Hubble time, from a low of about
4\% for his second and fourth models, to a high of about 17\% for his fifth model.

The cause of that evolution is that chaotic orbits change their behaviour with
time, resembling that of regular orbits at certain intervals, behaving more
chaotically at other intervals and exploring different regions of space in the
meantime. Moreover, that weaker or stronger chaotic behaviour can be traced with
Lyapunov exponents computed over finite intervals which decrease and increase
their values accordingly \cite{KM94}, \cite{Muz07}. Merritt and his coworkers
tried to solve this problem using what they called "fully mixed solutions"
\cite{MF96} and, more recently, integrating orbits over five Hubble times
\cite{Cap07}. In the former work, they found solutions for the weak cusp model,
but not for their strong cusp model; the subsequent evolution of these models to
test their stability was not investigated, however. In the latter work
they indicate that there is "a slight evolution toward a more prolate shape",
but they provide no quantitative estimates other than indicating that differences
in velocity dispersions are "almost always below 10\%". Clearly, it is very
difficult to incorporate chaotic orbits in Schwarzschild's method: as some
chaotic orbits begin to behave more chaotically, one needs to have other chaotic
orbits that behave more regularly as compensation; such a delicate equilibrium
cannot be attained simply obtaining the weights of chaotic orbits over longer
integration times and, moreover, the relatively low number of orbits used
(typically a few thousands) makes even more difficult that task. Finally, the
usual imposition of constant axial ratios over the whole system in
Schwarzschild's method prevents the existence of a rounder halo of chaotic
orbits that seems to be a necessary condition to have highly chaotic triaxial
stellar systems \cite{VKS02}, \cite{MCW05}, \cite{AMNZ07}.

\subsection{The stability of highly chaotic triaxial stellar systems}
\label{sec:5}
The models of the N--body method are built self--consistently from the
start and typically contain hundreds of thousands, or even millions, of bodies so
that they should be free of the difficulties that plage the construction of highly
chaotic triaxial stellar systems with Schwarzschild's method. In fact, stable
models with high fractions of chaotic orbits were obtained with the N-body method,
using about $10^5$ particles \cite{VKS02}, \cite{MCW05}; moreover, the fractions
of the different types of orbits were not significantly altered when the potential
was fitted to the N--body distribution at different times. A stable cuspy model
that was mildly triaxial and made up of 512,000 particles, plus several others with
128,000 particles, were also built \cite{HB01}; later on, it was shown that the
introduction of a black hole, although affecting the inner regions of the model,
did not alter the triaxiality at larger radii and the authors concluded that the
triaxiality of elliptical galaxies is not inconsistent with the presence of
supermassive black holes at their centers \cite{HB02}.

Highly stable models of $10^6$ particles were built by us with the N--body
method \cite{AMNZ07},
\cite{MNZ08}: all of them have decreasing flattening from center to border,
which arised naturally from the N--body evolution during the generation of the
systems; they have different degrees of flattening and triaxiality, two
of them are moderately cuspy ($\gamma \approx 1.0$), and all have high
fractions (between 36\% and 71\%) of chaotic orbits. When integrated with the
N--body code, our models suffer changes in their central density and minor
semiaxis values which do not exceed, respectively, about 4\% and 2\% over a
Hubble time. Nevertheless, these changes are most likely due to collisional
effects of the N--body code \cite{HB90} because, when the number of bodies is
reduced by a factor of 10 (and their masses are increased by the same factor), 
those changes increase by factors between 3 and 10. Alternatively, integrating
the motion of the bodies in the fixed smooth potential, which suppresses the
collisional effects (and which, by the way, is what Schwarzschild did) reduces
those changes to 0.1\% only (i.e., between one and two orders of magnitude smaller
than those found by Schwarzchild \cite{Sch93}).

Thus, we may conclude that highly stable triaxial models with large fractions
of chaotic orbits can be built with the N--body method. The difficulties to
build such models with Schwarzschild's method should thus be attributed to the
method itself and not to physical reasons.

\section{Partially and fully chaotic orbits}
\label{sec:6}
Since we are dealing with stationary systems, the orbits of the particles that
make them up obey the
energy integral, but they need two additional isolating integrals to be regular
orbits. Thus, we distinguish between partially chaotic orbits (one additional
integral besides energy) and fully chaotic orbits (energy is the only integral
they obey). A practical way to make the distinction is to compute the six
Lyapunov exponents: they come in three pairs of equal value and opposite signs,
due to the conservation of phase space volume, and each isolating integral
makes zero one pair. Thus, in our case, two Lyapunov exponents are always
zero (due to energy conservation); of the remaining four, if two are positive
the orbit is fully chaotic, if only one is positive the orbit is partially
chaotic and, finally, if all are zero the orbit is regular.

It was noted in \cite{PV84} that orbits obeying two isolating integrals have
smaller fractal dimension than orbits obeying only one, but earlier hints of
the differences between them can also be found in \cite{GS81} (whose
semi-stochastic orbits are probably what we now call partially chaotic orbits)
and in \cite{CGG78} (whose orbits in their big and small seas can be identified,
respectively, with the fully and partially chaotic orbits).

The reason why distinguishing partially from fully chaotic orbits in galactic
dynamics is important is that, since they obey different numbers of isolating
integrals, they have different spatial distributions as shown in \cite{Muz03}, 
\cite{MM04}, \cite{MCW05}, \cite{AMNZ07} and \cite{MNZ08}. In triaxial systems,
partially chaotic orbits usually exhibit a distribution intermediate between
those of regular and of fully chaotic orbits, and a possible explanation is that
some of the partially chaotic orbits lie in the stochastic layer surrounding
the resonances and thus behave similarly to regular orbits \cite{NPM07}.
Nevertheless, that is not the whole story as some partially chaotic orbits
seem to obey a global integral, rather than local ones \cite{AMNZ07}.

Partially chaotic orbits should not be confused with fully chaotic orbits with
low Lyapunov exponents, which also tend to have distributions more similar to
those of regular orbits than those of fully chaotic orbits with high Lyapunov
exponents \cite{MM04}, \cite{MCW05}. It is worth recalling that, no matter how
small their Lyapunov exponents are, fully chaotic orbits obey only one
isolating integral while partially chaotic orbits obey two so that, from a
theoretical point of view, they are indeed different kinds of orbits. From a
practical point of view, it is also easy to see that they have different
distributions: Table 1 gives the axial ratios of the distributions of
different kinds of orbits for models E4, E5 and E6 from \cite{AMNZ07} and E4c
and E6c from \cite{MNZ08}; the x, y and z axes are parallel, respectively, to
the major, intermediate and minor axes of the models. The third column gives
the axial ratios for the distributions of partially chaotic orbits, and the
fourth and fifth columns give the same ratios for weakly fully chaotic orbits
for two choices of the limiting value of the Lyapunov exponents used to
define "weakly", 0.050 and 0.100. Although for some models (e.g. E4 and E4c)
the possible differences are masked by the rather large statistical errors,
it is clear from the Table that the distributions of partially chaotic orbits
are significantly different from those of weakly fully chaotic orbits (at the
$3 \sigma$ level) for the other models.

\begin{table}
\centering
\caption{Axial ratios of the different classes of orbits in our models.}
\label{tab:1} 
\begin{tabular}{lllll}
\hline\noalign{\smallskip} 
Ratio & ~~System & ~~Partially Ch. & ~~~W.F.Ch. (0.050) & ~~W.F.Ch. (0.100)\\ 
\noalign{\smallskip}\hline\noalign{\smallskip}
y/x & ~~E4 & ~~$0.896 \pm 0.064$ & ~~~$0.692 \pm 0.027$ & ~~$0.745 \pm 0.019$ \\ 
 & ~~E5 & ~~$0.808 \pm 0.036$ & ~~~$0.764 \pm 0.024$ & ~~$0.797 \pm 0.017$ \\ 
 & ~~E6 & ~~$0.658 \pm 0.035$ & ~~~$0.789 \pm 0.027$ & ~~$0.845 \pm 0.019$ \\
 & ~~E4c & ~~$0.748 \pm 0.027$ & ~~~$0.733 \pm 0.016$ & ~~$0.730 \pm 0.013$ \\ 
 & ~~E6c & ~~$0.528 \pm 0.020$ & ~~~$0.693 \pm 0.013$ & ~~$0.700 \pm 0.010$ \\
\noalign{\smallskip}\hline\noalign{\smallskip}
z/x & ~~E4 & ~~$0.790 \pm 0.054$ & ~~~$0.802 \pm 0.035$ & ~~$0.826 \pm 0.024$ \\ 
 & ~~E5 & ~~$0.477 \pm 0.018$ & ~~~$0.684 \pm 0.021$ & ~~$0.708 \pm 0.014$ \\ 
 & ~~E6 & ~~$0.286 \pm 0.013$ & ~~~$0.644 \pm 0.022$ & ~~$0.673 \pm 0.015$ \\
 & ~~E4c & ~~$0.692 \pm 0.024$ & ~~~$0.762 \pm 0.017$ & ~~$0.757 \pm 0.013$ \\ 
 & ~~E6c & ~~$0.334 \pm 0.010$ & ~~~$0.466 \pm 0.007$ & ~~$0.490 \pm 0.006$ \\
\noalign{\smallskip}\hline
\end{tabular} 
\end{table} 

At any rate, it is clear that the distributions of partially and fully chaotic
orbits differ significantly and that they should not be bunched together as a
single group of chaotic orbits. The problem is that the computation of the
Lyapunov exponents demands long computation times and there are not yet faster
methods that allow to distinguish partially from fully chaotic orbits. The fact
that many chaotic orbits can be frequency analyzed and are found to lie in
regions of the frequency map corresponding to regular orbits \cite{KV05} might,
perhaps, lead to a faster method of separation in the future. Nevertheless, many
fully chaotic orbits can be frequency analyzed, while many partially chaotic
orbits cannot \cite{AMNZ07}, so that much remains to be done before a workable
method based on frequency analysis can be designed.

\section{Figure rotation in triaxial systems}
\label{sec:7}
Although the system investigated in \cite{MCW05} had been regarded as stationary,
integrations much longer than those used in that work revealed that, in fact, it
was very slowly rotating around its minor axis \cite{Muz06}. The total angular
momentum of the system was zero, so that this was an unequivocal case of figure
rotation. 

Figure rotation was also found in most of the models studied in \cite{AMNZ07} and
\cite{MNZ08} and it is clear that the rotational velocity increases with the
flattening of the system; only model E4 from \cite{AMNZ07}, which is almost axially
symmetric, prolate and with axial ratio close to 0.6 has no significant rotation. It
should be stressed, however, that even the highest rotational velocities found thus
far are extremely low: the systems can complete only a fraction of a revolution in a
Hubble time or, put in a different way, the radii of the Lindblad and corotation
resonances are at least an order of magnitude larger than the systems themselves.

It had been suggested that figure rotation might produce important changes in the
degree of chaoticity \cite{MF96} and it turned out that, in spite of the extremely
low rotational velocity, a significant difference in the fraction of chaotic orbits
was found between the models of \cite{MCW05} and \cite{Muz06} which only differ in
that the former is stationary and the latter is rotating. Alternatively, no
significant difference was found for the different kinds of regular orbits in those
two models. The most likely explanation is that, although the rotational velocity
is too low to produce a measurable effect on the regular orbits, the break of
symmetry caused by the presence of rotation suffices to increase chaos significantly.

\section{Musings on orbital classification through frequency analysis}
\label{sec:8}

\subsection{Classification methods}
\label{sec:9}
The spectral properties of galactic orbits were investigated by Binney and Spergel
\cite{BS82} and, more recently, Papaphilippou and Laskar \cite{PL96} and
\cite{PL98} applied to stellar systems the frequency analysis techniques developed
by the latter for celestial mechanics. Following the ideas of Binney and Spergel,
Carpintero and Aguilar \cite{CA98} developed an automatic orbit classification code.
Kalapotharakos and Voglis \cite{KV05} developed a classification system based on
the frequency map of Laskar and, later on, I \cite{Muz06} improved it somewhat.

Having used extensively both the Carpintero and Aguilar \cite{CMW99}, 
\cite{MCW00}, \cite{CVM01}, \cite{CMVW03} and \cite{MCW05}, and the Kalapotharakos
and Voglis methods \cite{Muz06}, \cite{AMNZ07} and \cite{MNZ08}, I strongly prefer
the latter. The main advantage of the Kalapotharakos and Voglis method is that one
can see what is happening throughout the process. It is very easy to detect problems
from the anomalous positions that the corresponding frequency ratios yield on the
frequency map and, thus, to improve the method. This is an aspect that deserves to
be emphasized: the need to use frequencies different from those corresponding to the
maximum amplitudes had not been noted in \cite{KV05}, but it was in \cite{Muz06},
probably because a somewhat cuspier potential was investigated in the latter
work; similarly, that distinction was unnecessary for the long axis tubes (LATs
hereafter) of \cite{Muz06}, but had to be made for those of the almost axially
symmetric E4 system of \cite{AMNZ07}. In other words, as one explores different
stellar system models (cuspier, closer to axisymmetry, and so on)
the orbital classification system may need to be improved and that need is quite
evident with the Kalapotharakos and Voglis method. Thanks to these improvements,
virtually all the regular orbits can be classified with the frequency map,
while usually between 10\% and 15\% of them remain unclassified with the other
method \cite{MCW05} and \cite{J05}. 

Besides, separation of chaotic from regular
orbits with the method of Carpintero and Aguilar is erratic, at least in rotating
systems \cite{CMVW03}. Since the problem seems to arise from the presence of nearby
lines in the spectra, which is worse in rotating systems but not exclusive of them,
I strongly suspect that orbit classification in non--rotating systems may also be
affected. That is why in our last work with that method \cite{MCW05} we used
it only to classify regular orbits, previously selected using Lyapunov exponents.

\subsection{Which frequency to choose?}
\label{sec:10}
Frequency analysis is usually performed on complex variables formed taking one
coordinate as the real part and the corresponding velocity as the imaginary part.
One thus gets the frequencies Fx, Fy and Fz corresponding, respectively, to motion
along the (x, y, z) axes which, in turn, are parallel to the main axes of the
stellar system. The frequencies usually selected for the frequency map are those
corresponding to the maximum amplitudes in each coordinate \cite{WFM98}, \cite{KV05},
but it has been known since 1982 \cite{BS82} that, due to a libration effect, one
should not always take those. Besides, another effect linked to very highly elongated
orbits also demands to adopt frequencies which are not the ones corresponding to the
maximum amplitudes \cite{Muz06}, \cite{AMNZ07}. Nevertheless, it is just fair to note
that these exceptions are not too common: out of 17,103 orbits investigated in
\cite{AMNZ07} and \cite{MNZ08} only 265 (1.5\%) needed the former correction and 153
(0.9\%) the latter one. These fractions vary considerably from one model to another,
however, and as the affected orbits tend to concentrate at low absolute values of
energy and/or are extremely elongated, not taking these effects into account
might bias the sample of classified orbits.

\subsection{The usefulness of the energy vs. frequency plane}
\label{sec:11}
Regular orbits obey two additional isolating integrals, besides energy, and the
values of the orbital frequencies are related to these integrals. For a given
energy, different frequencies imply different values of the other integrals and,
thus, different types of orbits. Inner and outer LATs,
short axis tubes, boxes and even different resonant orbits can be separated on the
energy vs. frequency (or frequency ratio) plane, but that does not mean that it is
practical to use it, because those separations are more easily done on the
frequency map.

Nevertheless, some insight can be gained from the use of the energy vs.
frequency plane. Figure 1 of \cite{Muz06} offers a good example, because that
plane was used there to show that one should not always use the frequency
corresponding to the maximum amplitude as the principal frequency. Besides,
while in \cite{KV05} it was correctly stated that outer LATs had larger
Fx/Fz values than inner LATs, no indication of which was the separating
value was provided there. Actually, as shown in Figure 2 of \cite{AMNZ07}, one has
to use the energy vs. frequency ratio plane to separate inner from outer LATs,
because the separating value varies with the energy of the orbit.

\begin{figure}
\vskip 5.5mm
\centering
\includegraphics[width=5.5truecm]{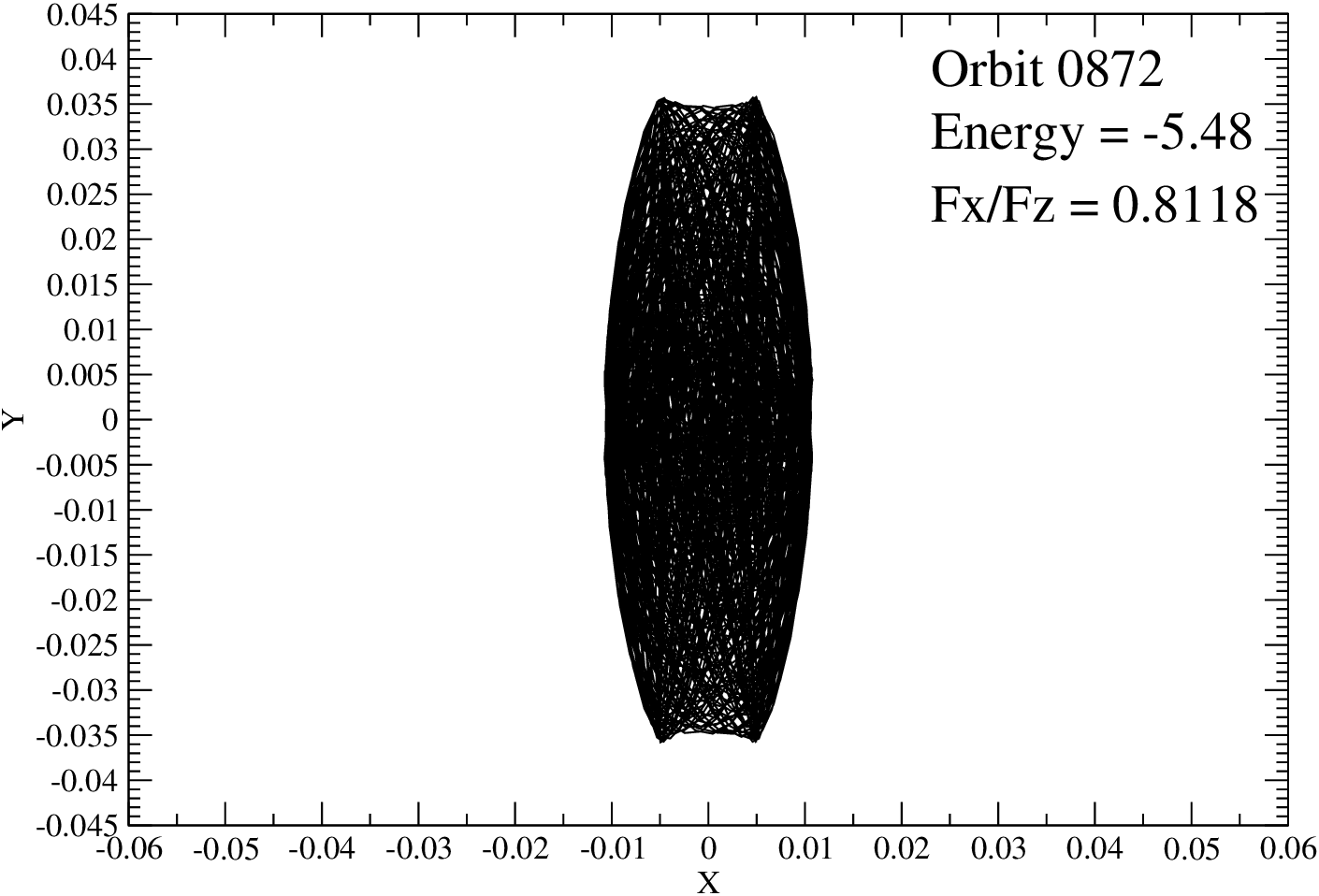}~\hfill
\includegraphics[width=5.5truecm]{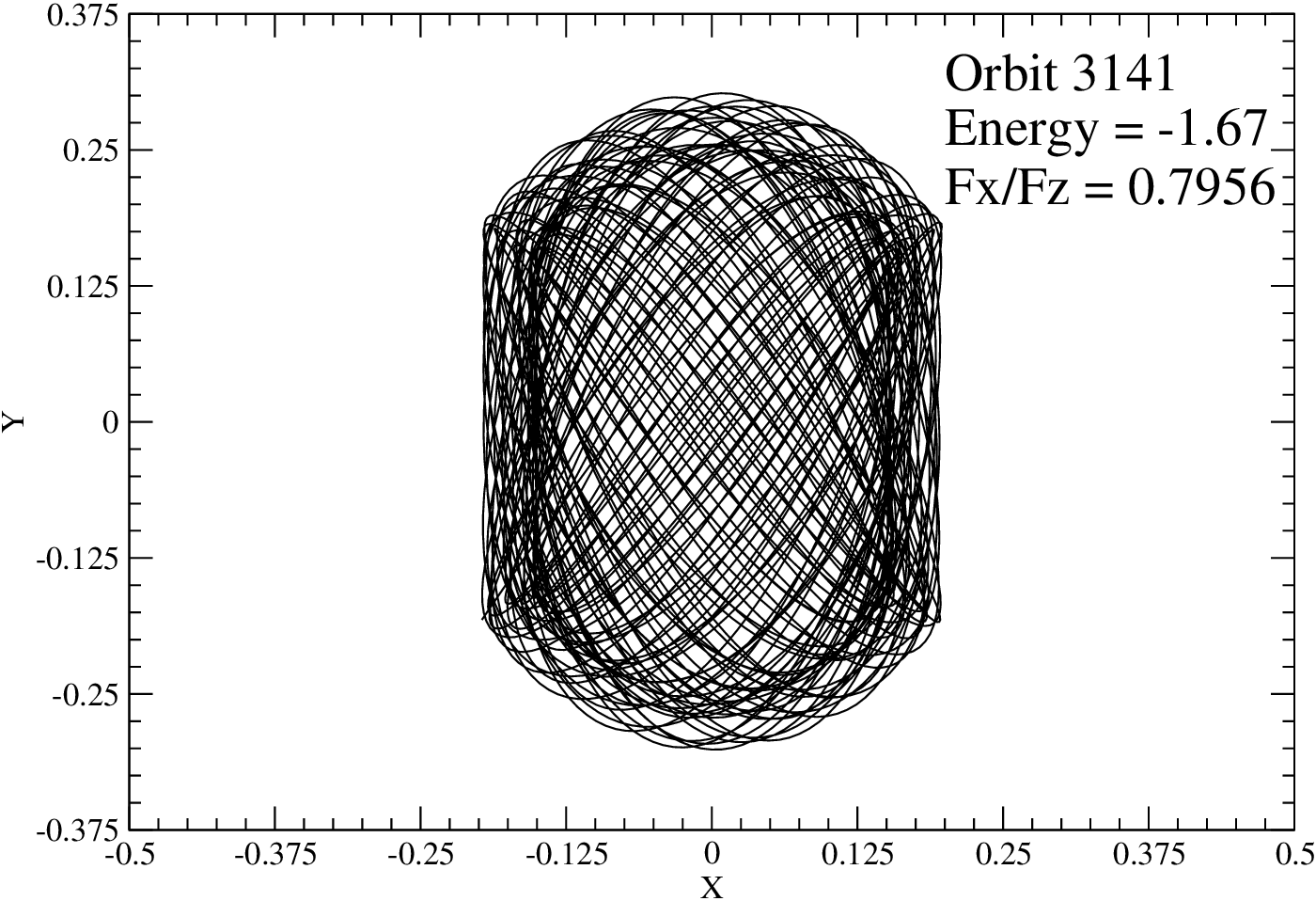}
\vskip 6.5mm
\centering
\includegraphics[width=5.5truecm]{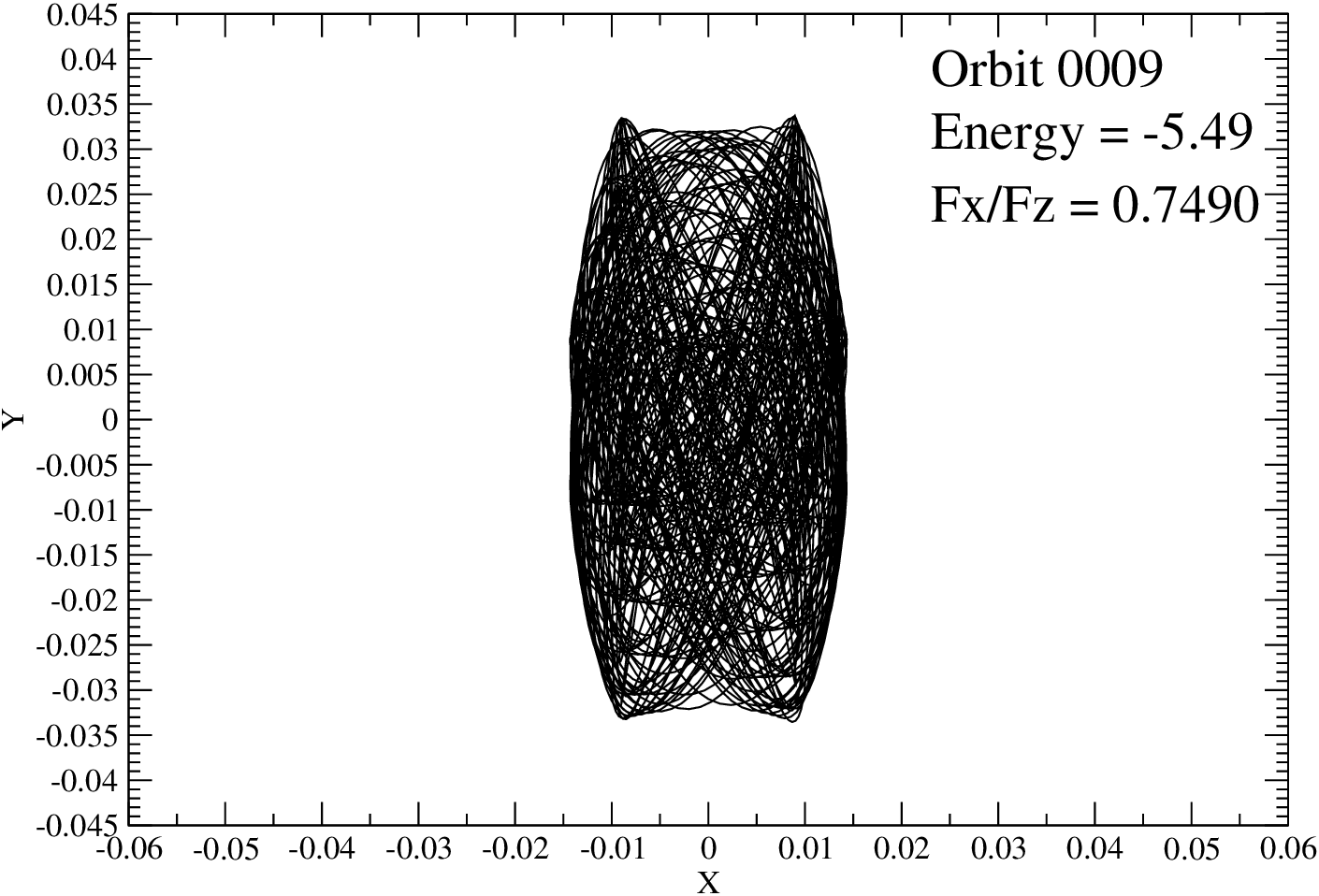}~\hfill
\includegraphics[width=5.5truecm]{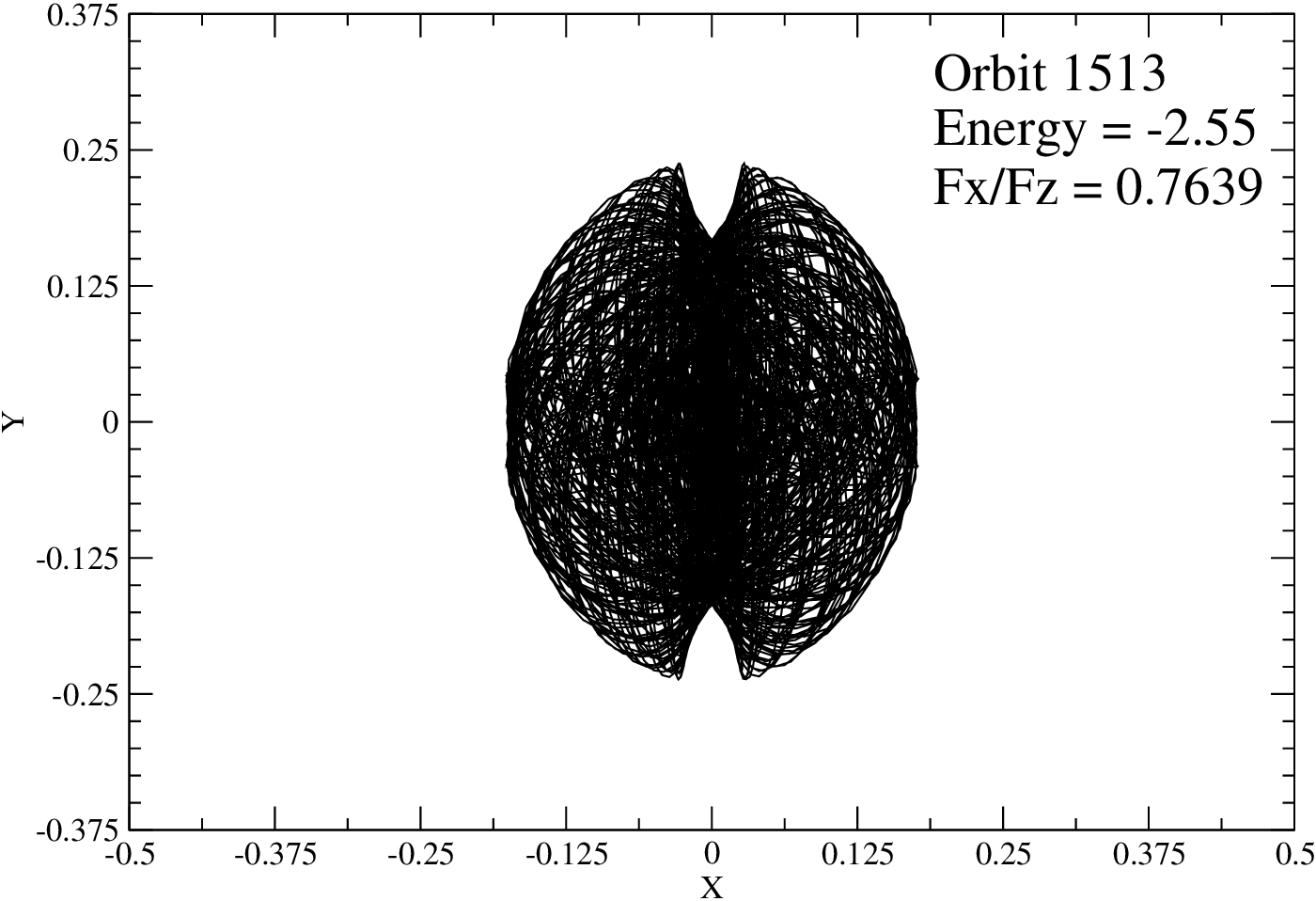}
\vskip 6.5mm
\centering
\includegraphics[width=5.5truecm]{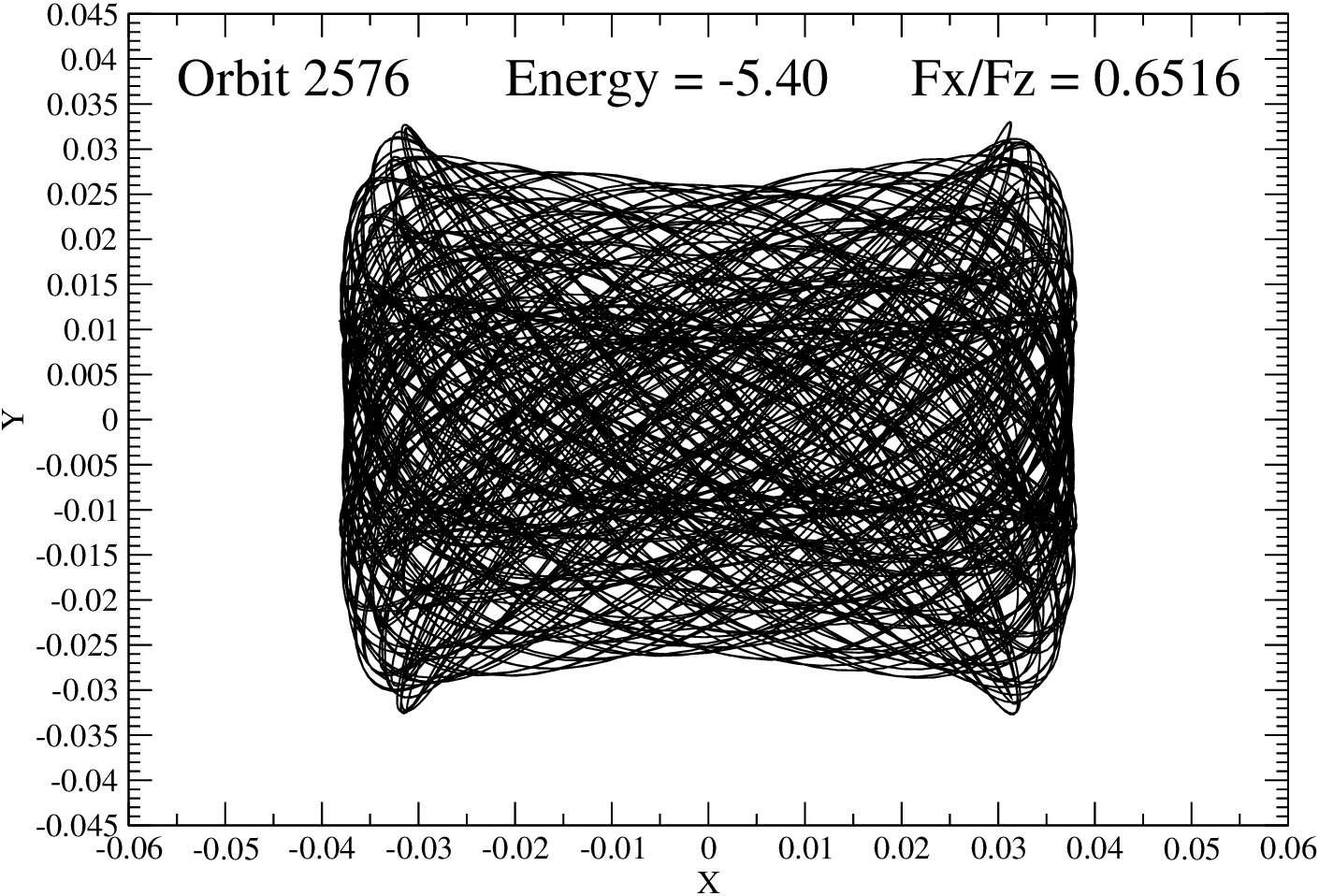}~\hfill
\includegraphics[width=5.5truecm]{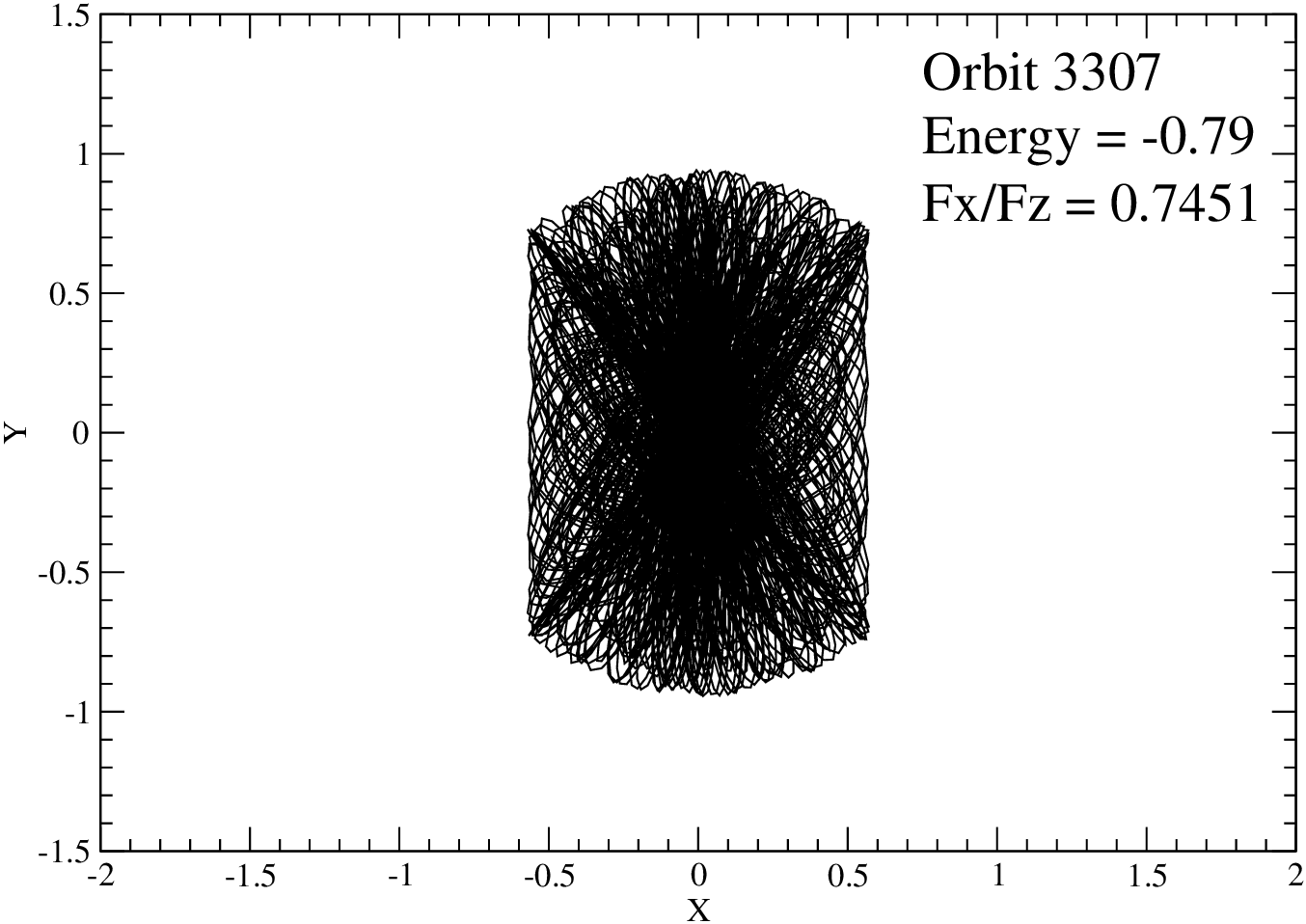}
\caption{Projection on the (x, y) plane of several examples of LATs; see text for details}
\label{fig:1}
\end{figure}

Figure 1 presents the (x, y) projections of several LATs from model
E4 of \cite{AMNZ07}. The orbits on the left column have similar energy values,
close to the minimum energy of -5.96 and, although their Fx/Fz values range from
0.6516 to 0.8118, they are all inner LATs, as evidenced by their concave
upper and lower limits. We also notice that their extension along the x axis is
reduced as their Fx/Fz values increase and, in fact, the regions of space occupied
by orbits 0872 and 0009 resemble more those occupied by outer LATs than
those occupied by inner LATs. We found a similar effect on the x extension
of the orbits at other energy values although, when the separation shown in Figure 2
of \cite{AMNZ07} is crossed, there is also of course a change from inner to outer
LATs. The upper and middle parts of the right column of Figure 1 correspond
to orbits 3141 and 1513 that are virtually face to face at each side of the separation
on the energy vs. frequency ratio plot: they have similar Fx/Fz values but, due to their
energy difference, the former is an outer, and the latter an inner, LAT. Notice also that
the Fx/Fz value of (outer LAT) orbit 3141 is lower than that of (inner LAT) orbit 0872.
Finally, the lower right section of Figure 1 corresponds to (outer LAT) orbit 3307, whose
Fx/Fz value is lower than those of (inner LAT) orbits 0009 and 0872.

\begin{figure}
\vskip 5.5mm
\centering
\includegraphics[width=5.5truecm]{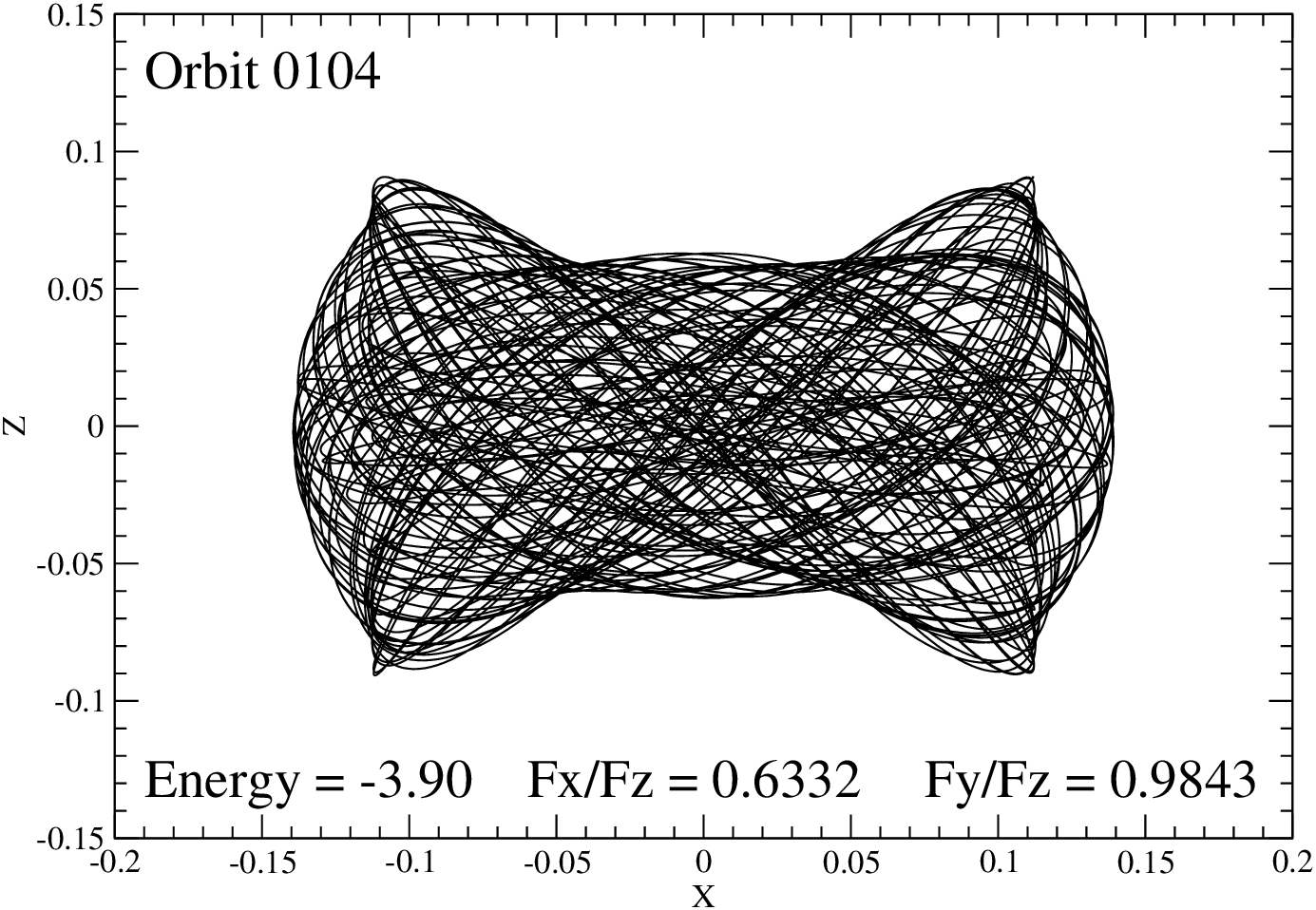}~\hfill
\includegraphics[width=5.5truecm]{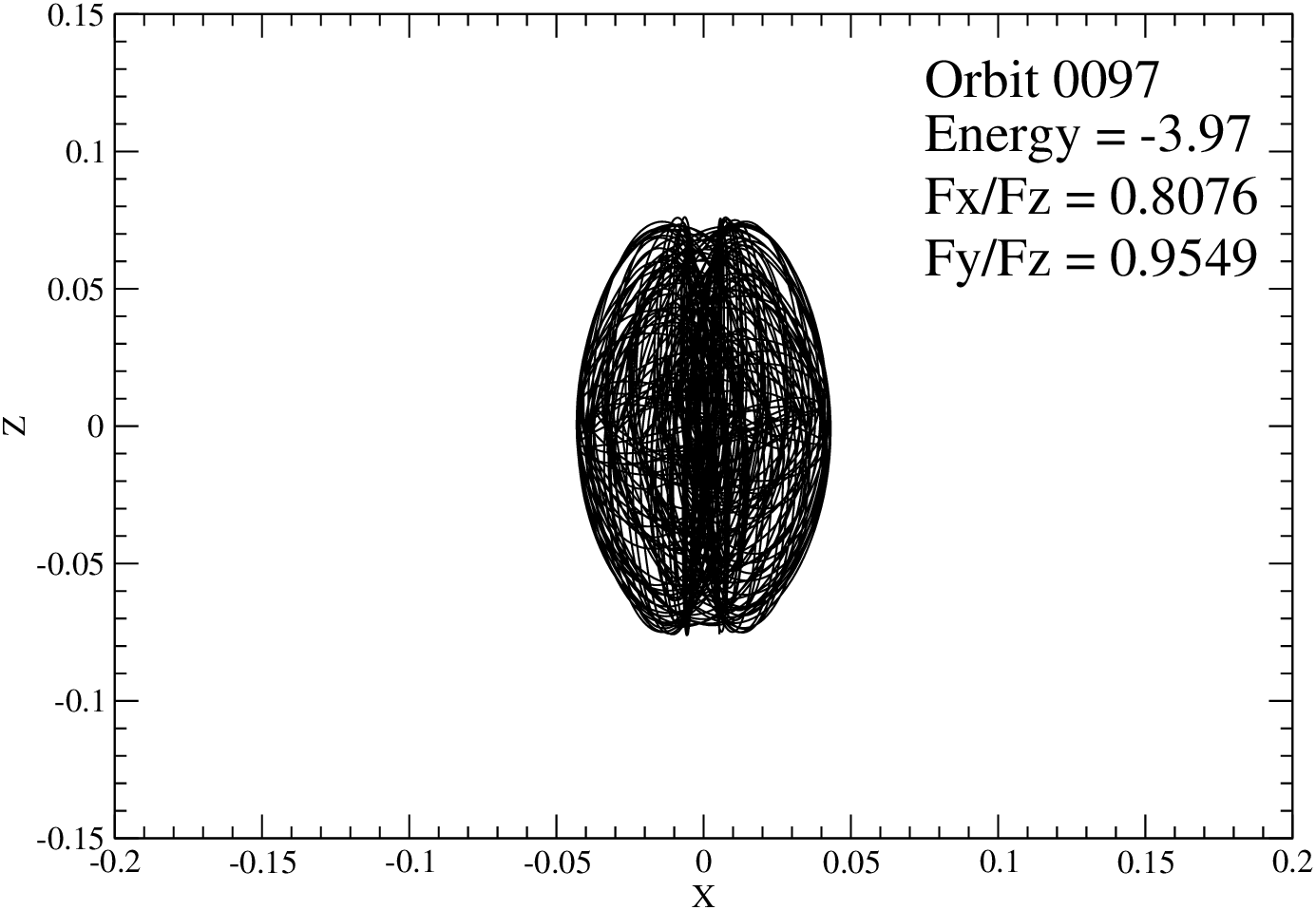}
\vskip 6.5mm
\centering
\includegraphics[width=5.5truecm]{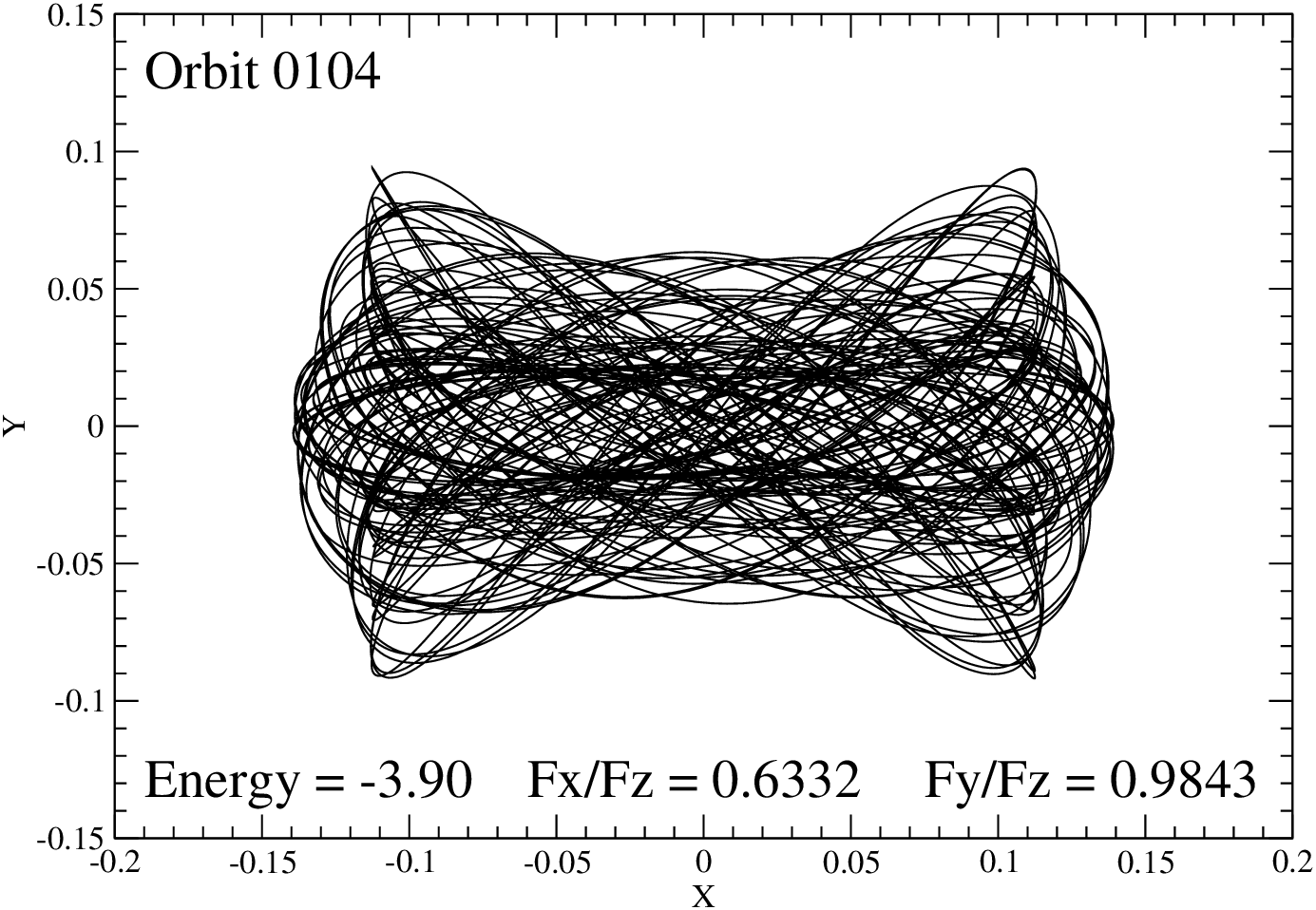}~\hfill
\includegraphics[width=5.5truecm]{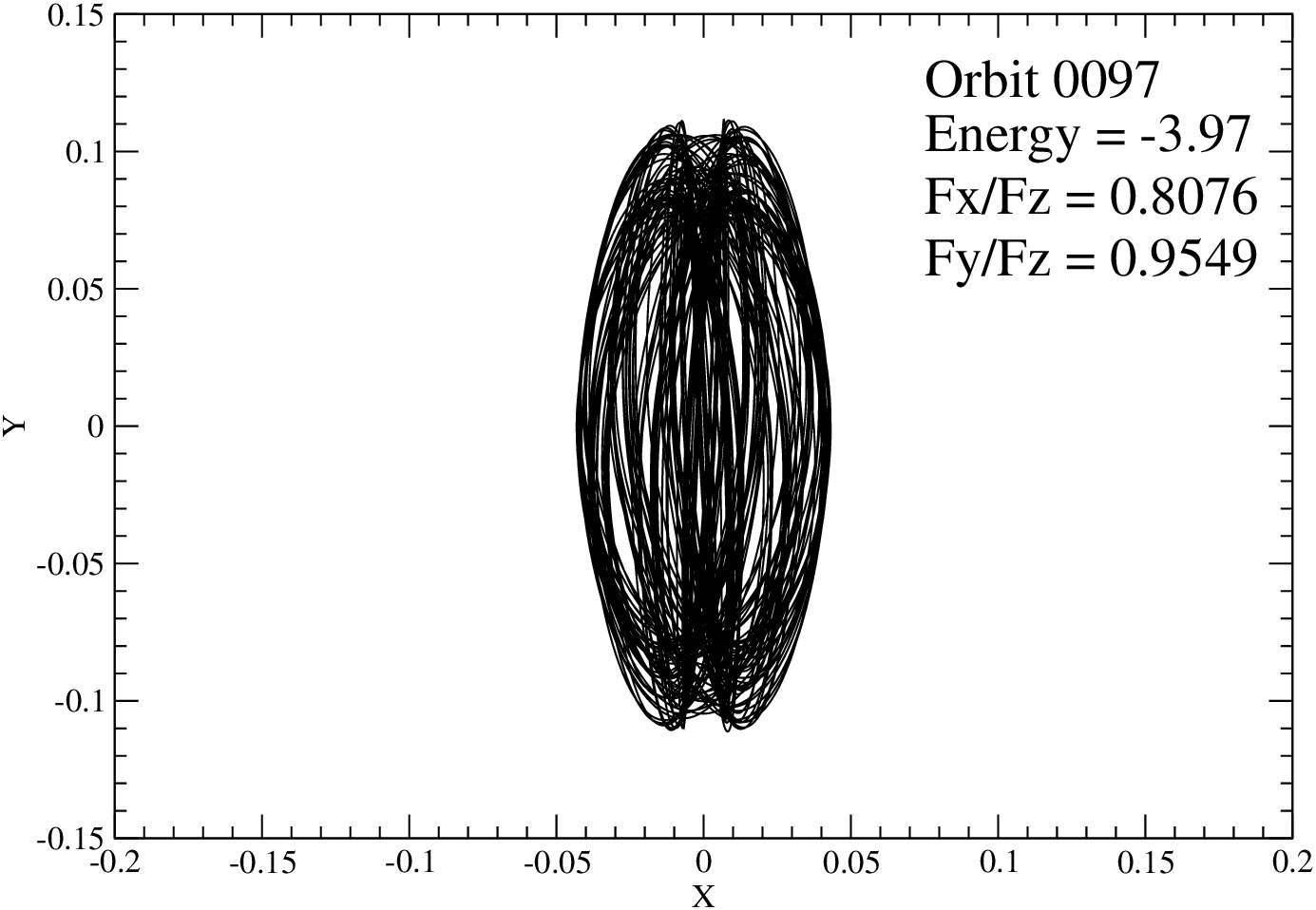}
\caption{Projection on the (x, y) plane of two examples of boxes; see text for details}
\label{fig:2}
\end{figure}

Interestingly, the shortening of the x axis as the Fx/Fz ratio increases, shown above for
the LATs, affects the boxes as well. Figure 2 presents the (x, y) and (x, z) projections of
orbits 0104 and 0097 from model E4 of \cite{AMNZ07}, which have both essentially the same
energy. Nevertheless, while the former, with Fx/Fz = 0.6332, lies straight on the line
occupied by the boxes on the energy vs. frequency ratio plane, the latter, with Fx/Fz =
0.8076, lies well above that line. We see on the left part of Figure 2 that 0104 is indeed
a typical box, but the right part shows that 0097, although still a box, is strongly
compressed along the x axis.

Due to their elongation along the major axis, inner LATs and boxes are usually considered
as the main building blocks of highly elongated triaxial systems, but we now see that
there are inner LATs and boxes that are, in fact, strongly compressed along that axis. To
put things in the proper perspective we should emphasize, however, that these orbits were
found in the almost rotationally symmetric model E4 of \cite{AMNZ07} and that they are not
very abundant. 

\section{Discussion}
\label{sec:12}
We have reviewed several papers on triaxial stellar systems built with the N--body method
that show that it is perfectly possible to have strongly chaotic triaxial stellar systems
that are also highly stable over periods of the order of a Hubble time. The difficulties
to build such systems with Schwarzschild's method should thus be attributed to the
method itself and not to physical reasons.

It is clear, both from a theoretical and from a practical point of view, that partially
and fully chaotic orbits populate different regions of space and should not be bunched
together under the single banner of chaotic orbits. The main problem here is that the
single method thus far available to separate them, that of Lyapunov exponents, is very
slow and faster methods are wanted. We also showed that the distribution of partially
chaotic orbits is different from that of weakly fully chaotic orbits, in accordance
with the fact that the former obey two isolating integrals of motion and the latter only
one.

Very slow figure rotation seems to be an ordinary trait of strongly elongated triaxial
stellar models formed through the collapse of cold N--body systems. The rotational velocity
diminishes, and even disappears entirely, as one goes to less elongated and less triaxial
models.

Frequency analysis offers a very useful tool for the classification of large numbers of
regular orbits. I strongly favor the use of the method of Ka\-la\-po\-tha\-ra\-kos and Voglis
\cite{KV05}, with the improvements we introduced in \cite{Muz06} and \cite{AMNZ07}. Since
the need for those improvements became apparent when models with different characteristics
(cuspiness, approximate rotational symmetry) were considered, it would not be surprising
that further refinements will be necessary as the method is applied to other systems.
Nevertheless, a nice feature of this method is that, when there is such need, it becomes
plainly evident. Besides, plots of known integrals, such as energy, and the orbital frequencies
(or frequency ratios), that are related to the values of the integrals, are very useful to
reveal peculiarities of the orbits as one explores different models; a good example of this
is provided by the compression along the major axis of some LATs and boxes from an almost
axisymmetric system, shown in our Figures 1 and 2.

\section{Acknowledgements}
\label{sec:13}
I am very grateful to H\'ector R. Viturro and to Ruben E. Mart\'{\i}nez for their
technical assistance, and to Lilia P. Bassino and to an anonimous referee for carefully
reading the first version of this paper and suggesting some language improvements.
This work was supported with grants from the Consejo Nacional de
Investigaciones Cient\'{\i}ficas y T\'ecnicas de la Rep\'ublica Argentina, the
Agencia Nacional de Promoci\'on Cient\'{\i}fica y Tecnol\'ogica and the Universidad
Nacional de La Plata.

\end{document}